\documentclass{iaus}
\usepackage{graphicx}
\usepackage{subfigure}

\title[The expected photometrical characteristics of high redshift
galaxies] 
{The expected photometrical characteristics of high redshift spiral galaxies}

\author[Moll\'{a} et al.]{M. Moll\'{a} $^{1}$, 
M. Garc\'{\i}a-Vargas $^{2}$ \and M. Mart\'{\i}n-Manj\'{o}n $^{3}$}

\affiliation{$^{1}$ Dpto. de Investigaci\'{o}n B\'{a}sica, CIEMAT,
Avda. Complutense 22, 28040, Madrid, (Spain) \\ 
email:{\tt mercedes.molla@ciemat.es}\\
[\affilskip]
$^{2}$ FRACTAL SLNE, C/ Tulip\'{a}n 2, p13, 1A , 28231 Las Rozas de Madrid,
(Spain)\\  email:{\tt marisa.garcia@fractal-es.com} \\
[\affilskip]
$^{3}$ Dpto. de F\'{\i}sica Te\'{o}rica, Universidad Aut\'onoma de
Madrid. 28049 Cantoblanco (Spain)\\
email:{\tt mariluz.martin@uam.es}
}

\pubyear{2009}
\volume{xxx}  
\pagerange{119--126}
\jname{Title of your IAU Symposium}
\editors{A.C. Editor, B.D. Editor \& C.E. Editor, eds.}
\begin{document}

\maketitle

\begin{abstract}
  The star formation and the subsequent metallic enrichment of 
spiral galaxies, occurring at the early phases of their evolution, produce
emission lines whose contribution may change the colors of these
objects. We have computed this contribution with consistent
chemo-spectro-photometrical theoretical models in order to calculate
the variations in the colors evolution.
\end{abstract}

\firstsection 

\section{Evolutionary and Photoionization Synthesis Models}

We present the PopStar evolutionary synthesis model in
\cite{mol09}(Paper I).  Selected isochrones are a new Padova set,
specifically computed for this piece of work with a broad age and
metallicity coverage, and a detailed treatment of mass-loss for both,
young (O, B, WR) and old ages (post-AGB until planetary nebula).  The
spectral energy distributions (SEDs) are calculated for each Single
Stellar Population (SSP) by including the nebular contribution.
Colors are calculated for these SEDs for different photometric
systems.  They result redder for the youngest stellar populations,
mainly for low Zmet, than those obtained by \cite{bc03} but they are
similar to those obtained by \cite{stb99} for ages younger than 1
Gyr. The old stellar populations show the same colors than
\cite{bc03}. Both results imply that our models are equally well tuned
for young as for old stellar populations.

We have then computed photoionization models (CLOUDY) with the
previously described SSP-SEDs, obtaining the emission line spectra due
to the youngest stellar populations \cite{PaperII}(Paper II). Some
intense emission lines fall in the broad band filters and therefore
their contribution must be included into the magnitudes.  Colors of
stellar clusters changes appreciably, such as we may see in Fig.~3
from \cite[][Paper III]{gv09}, in particular when the emission lines
proceed from very young stellar clusters but also when they are as old
as 10 Myr. In the color-color diagrams, when a burst of star formation
takes places, points go out of the stellar population region, falling
in a region impossible to reach in any other way.

\section{Spiral Evolution Models}

The same basic chemical evolution numerical model was applied to a
wide grid of theoretical galaxies with different total masses and
variable star formation efficiencies in \cite{mol05}.  This grid
reproduces well the observational data for local spiral
galaxies.  By using the resulting star formation, $\Psi(t)$, and metal
enrichment, Z(t), histories, we calculate the SED, $F_{\lambda}(t)$,
for each galaxy from the equation:

\begin{equation}
F_{\lambda}(t)=\int_{0}^{t} S_{\lambda}(\tau,Z(t'))\Psi(t')dt',
\label{Flujo}
\end{equation}

where $\tau=t-t'$ and $S_{\lambda}(\tau,Z)$ is the SED of each stellar
generation or SSP of a given age $\tau$ and a metallicity $Z$, taken
from Paper I spectra. Once these SEDs calculated, we compute
magnitudes and colors in the Johnson and SDSS systems. They reproduce
well the data of our local universe, demonstrating that the grid
of models is well calibrated.
 
\section{Results and Conclusions}

\begin{figure}
\begin{center}
\includegraphics[width=2.5in,angle=0]{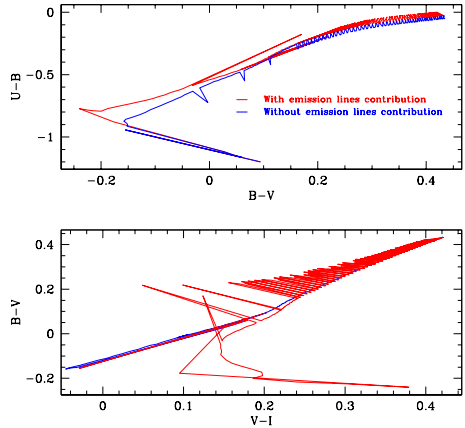}
\caption{Color-Color diagrams with and without the emission lines 
contribution}
\label{fig}
\end{center}
\end{figure}

We check if the observed colors of galaxies change when emission line
intensities are included in the calculations.  We compute the observed
and rest-frame colors evolution along the redshift for a MWG-type
galaxy with continuous star formation.  The rest-frame absolute colors
U-B and B-V evolution are shown in Fig.~\ref{fig}. The emission
lines produced by the last stellar generations change the magnitudes
in ~0.2 mag in any band.

In summary, the contribution of the emission lines when a burst takes
place changes the broad band magnitudes, therefore it is essential to
take it into account when star-forming galaxies are studied. This is
important specially at redshifts in which the star formation history
reaches its maximum. Therefore to interpreting high redshift colors by
using evolutionary synthesis models without taking into account the
star formation history of galaxies and the subsequent emission lines
may yield erroneous conclusions.

\end{document}